\documentclass[11pt]{article}

\usepackage[T1]{fontenc}
\usepackage[utf8]{inputenc}
\usepackage{lmodern}

\usepackage[margin=1in]{geometry}
\usepackage{microtype}
\usepackage{authblk}

\usepackage{amsmath,amssymb}

\usepackage{graphicx}

\usepackage{algorithm}
\usepackage{algpseudocode}

\usepackage{booktabs}
\usepackage{tabularx}
\usepackage{multirow}
\usepackage{makecell}

\usepackage{subcaption}

\usepackage{xcolor}

\usepackage{placeins}

\usepackage[colorlinks=true,
            linkcolor=blue,
            citecolor=blue,
            urlcolor=blue]{hyperref}

\begin{document}

\title{Stellar Age Compression Reshapes Interpretations of the Milky Way Thick-Disk Formation History}
\author[1,2]{Zhipeng Zhang}
\affil[1]{China Mobile Research Institute, Beijing 100053, China}
\affil[2]{China Mobile GBA (Greater Bay Area) Innovation Institute, Guangzhou 510656, China}
\date{\today}

\maketitle

\begin{center}
\textbf{Corresponding author:} Zhipeng Zhang (\texttt{zhangzhipeng@chinamobile.com})
\end{center}

\begin{abstract}
The formation timescale of the Milky Way thick disk is one of the central debates in Galactic archaeology.
The age--metallicity relation (AMR), formation timescale, and chemical evolution gradients are frequently used to infer a rapid assembly, short-timescale enrichment, and bursty formation history of the thick disk.
However, stellar ages are not directly observable, introducing the potential risk that inferred ages may harbor a systematic compression tied to observational quality.

In this paper, we use the same stellar sample and identical physical covariate matching conditions, but two independent age scales---spectroscopic inferred ages (astroNN) and asteroseismic ages (APOKASC-3)---to compare the observable signatures of the thick-disk formation history.
We find that several key observables previously supporting a rapid thick-disk formation are systematically weakened under seismic anchoring:
the AMR slope flattens from $-3.29$ to $-1.86\,\mathrm{Gyr\,dex^{-1}}$ ($\Delta a = +1.43$),
the formation timescale widens from $3.04$ to $3.55\,\mathrm{Gyr}$,
and the peak formation age shifts from $9.1$ to $6.0\,\mathrm{Gyr}$.

Through transport inversion experiments, we further show that additive noise can only broaden the age distribution and cannot reproduce the above pattern, whereas a compressive transport map ($\lambda < 1$) simultaneously reproduces a narrower age distribution, a steeper AMR, and rapid-formation-like observables.
This result indicates that the compression transformation itself is sufficient to generate rapid-formation-friendly observables without requiring an intrinsically bursty formation history.
Our findings reveal that statistical interpretations of the Milky Way formation history may depend sensitively on the stellar age definition itself.
\end{abstract}

\section{Introduction}

The formation timescale of the Milky Way thick disk is a key problem for distinguishing different disk evolution scenarios.
The age--metallicity relation (AMR) is widely used to infer the chemical enrichment rate, early enrichment history, and inside-out growth mode of the disk \cite{bovy2016}.
Some studies, based on spectroscopically inferred ages, propose that the thick disk formed rapidly within $\sim 1\,\mathrm{Gyr}$, exhibiting a bursty assembly mode supporting short-timescale chemical enrichment \cite{haywood2013,snaith2015}, while others argue for a more extended assembly \cite{xiang2022}.
This dispute directly relates to our fundamental understanding of the early evolutionary history of the Milky Way.

Stellar ages are not directly observable \cite{soderblom2010}.
Current methods estimate ages indirectly through spectroscopy, photometry, parallaxes, and data-driven pipelines (e.g., astroNN;~\cite{leung2019}); therefore, observational conditions (signal-to-noise ratio, parallax precision) may influence the final results through the inference process.
To obtain an independent benchmark, we introduce APOKASC-3 asteroseismic ages \cite{pinsonneault2025}, whose systematic errors arise from stellar oscillation frequencies and are thus distinct from those of spectroscopic inference, serving as a relatively independent external anchoring scale.

Recent studies indicate that spectroscopically inferred ages may exhibit a systematic compression dependent on observational quality: stars with lower signal-to-noise ratios or poorer parallax precision tend to have inferred ages that shrink toward the sample mean.
If such a compression effect exists, it will directly affect population statistics such as the width of the age distribution, the location of the peak, and the AMR slope, thereby potentially altering the physical interpretation of the thick-disk formation history.

The core question of this paper is not ``whether a bias exists,'' but rather: \textbf{Do the key observables of the Milky Way formation history themselves depend on the age definition?}
Using an identical stellar sample and the same physical covariate matching, we change only the age scale (inferred vs.\ seismic) and systematically compare diagnostic quantities including the AMR slope, formation timescale, peak formation age, and old-star fraction.
Through transport inversion experiments, we distinguish the compression effect from additive noise, thus revealing how an observationally driven compression reshapes the interpretation of the Milky Way formation history.

\section{Data and Methods}

\subsection{Data and Independent Age Scales}

The analysis sample is drawn from APOGEE spectroscopic survey data combined with Gaia parallaxes, together with astroNN inferred ages \cite{leung2019,mackereth2019} and APOKASC-3 asteroseismic ages \cite{pinsonneault2025}.
The two age scales have different systematic error origins:
\begin{itemize}
    \item \textbf{Inferred ages (astroNN)}: based on spectroscopy, photometry, and parallaxes, obtained through neural-network training \cite{leung2019}, and sensitive to observational quality;
    \item \textbf{Seismic ages (APOKASC-3)}: based on stellar oscillation frequencies \cite{pinsonneault2025,martig2016}, with a different systematic error budget, serving as an external anchoring scale rather than absolute truth.
\end{itemize}

We apply basic quality cuts: signal-to-noise ratio $\mathrm{SNR} > 50$, effective temperature $4000 < T_{\mathrm{eff}} < 5500\,\mathrm{K}$, surface gravity $2.0 < \log g < 3.5$, and parallax precision $\varpi/\sigma_{\varpi} > 5$.

\subsection{Coarsened Exact Matching: Ensuring Same-Sample Comparisons}

To exclude the influence of sample composition differences, we employ coarsened exact matching (CEM;~\cite{iacus2012}) to construct control samples in a five-dimensional physical covariate space: $T_{\mathrm{eff}}$, $\log g$, $[\mathrm{Fe/H}]$, $[\alpha/\mathrm{Fe}]$, distance, and Galactic coordinates $(R, Z)$.
During the matching process, \textbf{age information is excluded by construction}, ensuring that we compare the behavior of the same stars under different age definitions, rather than differences between distinct samples.

\subsection{Definition of the Compressive Transformation}

We introduce a linear transport/compression operator to describe the systematic relationship between inferred and seismic ages:
\begin{equation}
a_{\text{infer}} = \alpha + \lambda \cdot a_{\text{seismo}} + \epsilon,
\end{equation}
where $\lambda$ is the compression coefficient.
When $\lambda = 1$, the two age scales differ only by an additive offset; when $\lambda < 1$, the inferred ages systematically compress the age span---old stars are estimated younger, and young stars older---directly reducing the overall variance of the age distribution.

By contrast, an additive noise model $a_{\text{noisy}} = a_{\text{truth}} + \eta$ satisfies $\mathrm{Var}(X+\eta) \geq \mathrm{Var}(X)$, and can only broaden the distribution, not compress it.
This property provides the theoretical closure for the subsequent transport inversion experiments.
The key insight is: \textbf{only the compressive transformation can simultaneously narrow the age distribution and steepen the AMR}.

\subsection{Formation-History Diagnostics}

To quantify the impact on the interpretation of the Milky Way formation history, we define the following diagnostics and their astrophysical interpretations:

\begin{table}[htbp]
\centering
\caption{Formation-history diagnostics and their astrophysical interpretations.}
\label{tab:observables_def}
\begin{tabular}{ll}
\toprule
Diagnostic & Astrophysical interpretation \\
\midrule
Narrow age distribution (small $\Delta t$) & Rapid assembly \\
Old-dominated peak age & Early-dominated buildup \\
Steep negative AMR slope & Rapid enrichment \\
Early cumulative formation fraction (CFF) & Bursty formation history \\
\bottomrule
\end{tabular}
\end{table}

\subsection{Significance Estimation}

The statistical significance of all differences is estimated via bootstrap resampling ($N=10^4$), with the significance metric defined as $z = \Delta / \sigma_{\Delta}$.

\section{Results}

\subsection{Observationally Dependent Compression: Evidence for Compression}

\begin{figure}[htbp]
\centering
\includegraphics[width=\linewidth]{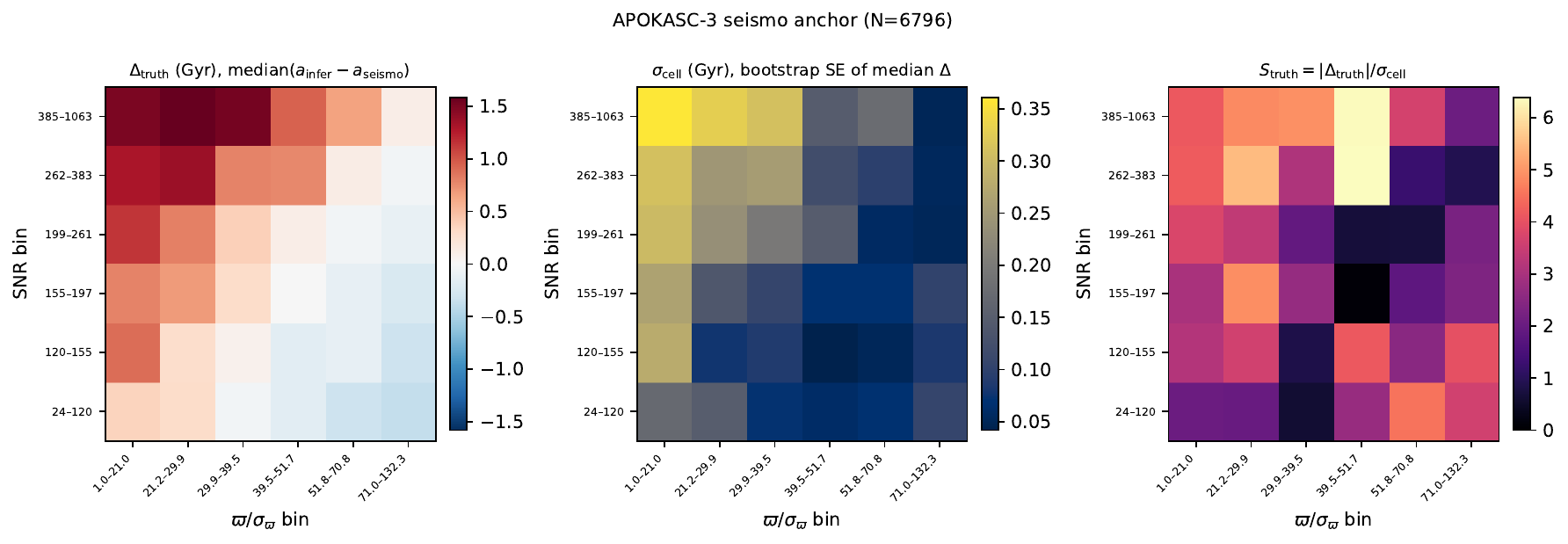}
\caption{Two-dimensional distribution of the inference bias anchored to APOKASC-3 seismic ages.
Left: median bias $\Delta_{\text{truth}}$;
Middle: bootstrap standard error $\sigma_{\text{cell}}$;
Right: significance $S_{\text{truth}} = |\Delta_{\text{truth}}|/\sigma_{\text{cell}}$.
Low-observational-quality regions exhibit significant structured biases.}
\label{fig:struth_basic}
\end{figure}

\begin{figure}[htbp]
\centering
\includegraphics[width=\linewidth]{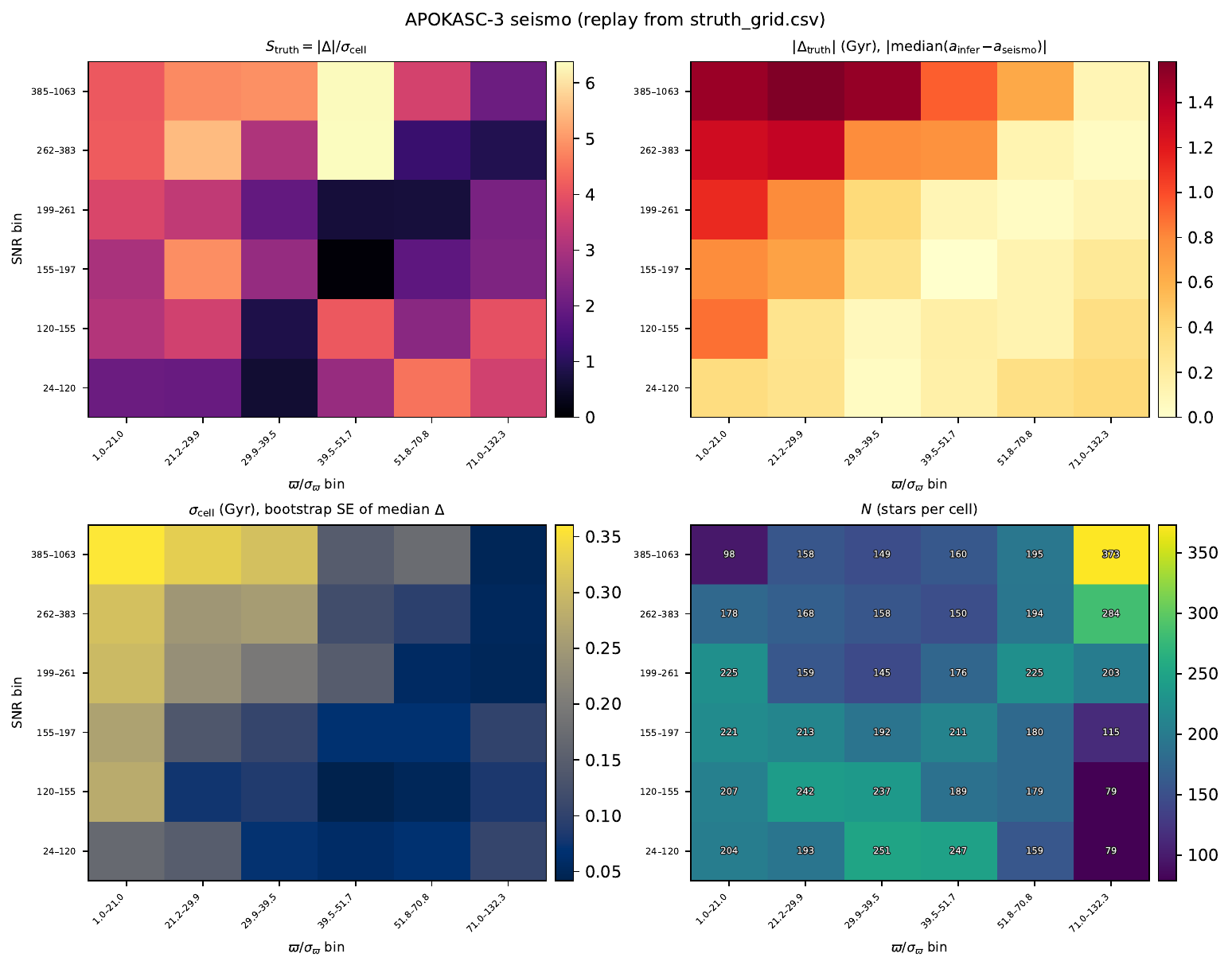}
\caption{Decomposition analysis for the same sample.
Top-left: $S_{\text{truth}}$;
Top-right: $|\Delta_{\text{truth}}|$;
Bottom-left: $\sigma_{\text{cell}}$;
Bottom-right: sample size $N$.
High $S_{\text{truth}}$ is primarily driven by systematic bias.}
\label{fig:struth_decomp}
\end{figure}

We first examine the distribution of the age difference between inferred and seismic ages in the observational parameter space.
Figure~\ref{fig:struth_basic} shows that in the two-dimensional bins of $\mathrm{SNR}$ and $\varpi/\sigma_{\varpi}$, the inferred bias $\Delta = a_{\text{infer}} - a_{\text{seismo}}$ exhibits a pronounced structured pattern: in low-observational-quality regimes, the bias amplitude reaches $\sim 0.5$--$1\,\mathrm{Gyr}$.
Figure~\ref{fig:struth_decomp} further decomposes this significance, showing that high-$S_{\text{truth}}$ regions are primarily driven by systematic bias rather than by statistical uncertainty or small sample size artifacts.

These observations demonstrate that the inferred bias is not random noise but a structured function of the observing conditions---indicating that compression indeed exists.
We next test whether such compression can alter the observed features of the Milky Way formation history.

\subsection{AMR Structure Under Different Age Definitions for the Same Sample}

On the strictly matched physical covariate sample, we compare the AMR structures under the two independent age scales.
Both panels use the identical stellar sample and the same $[\mathrm{Fe/H}]$ binning scheme, differing only in the age definition.

\begin{figure*}[t]
\centering
\includegraphics[width=0.9\textwidth]{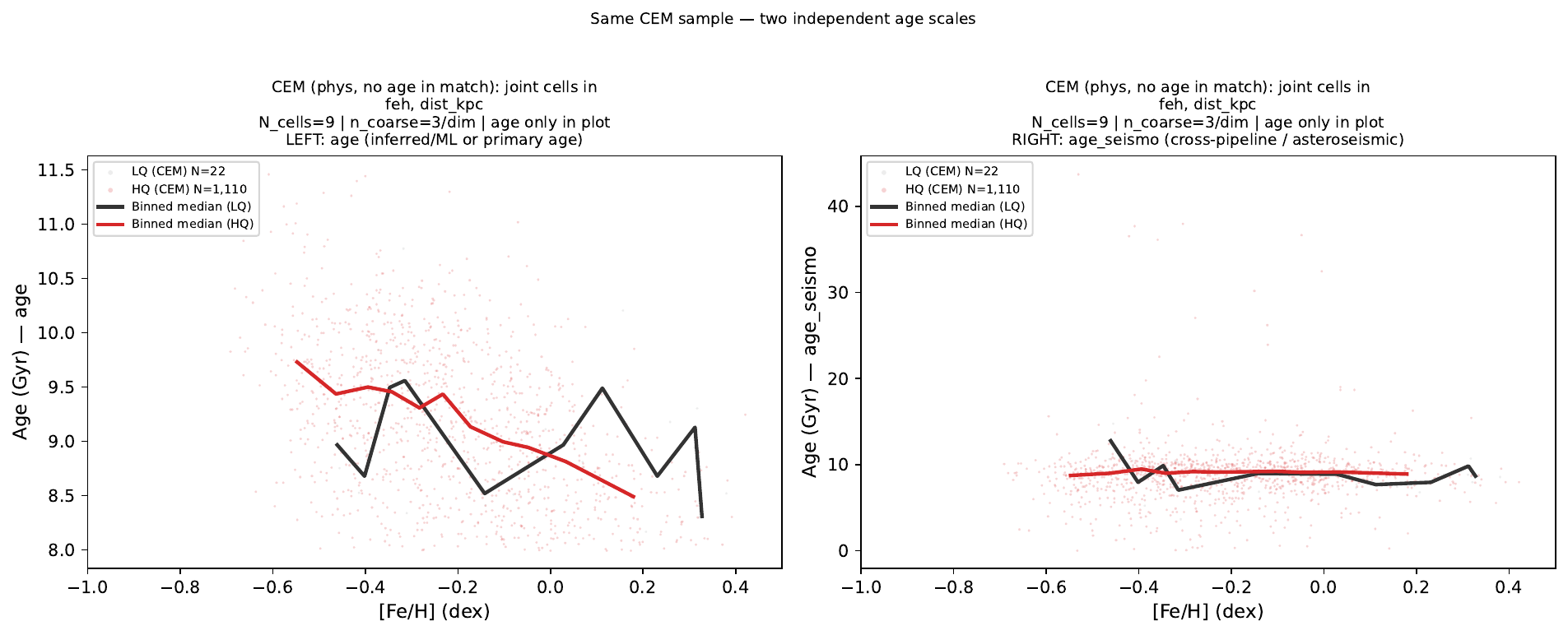}
\caption{AMR comparison from the same physically matched sample under two independent age scales.
Left: inferred ages (astroNN);
Right: seismic ages (APOKASC-3).
Both panels use the identical sample and binning; only the age definition differs.
Under seismic anchoring, the AMR flattens significantly.}
\label{fig:cem_twin_infer_seismo}
\end{figure*}

As shown in Fig.~\ref{fig:cem_twin_infer_seismo}, the AMRs under the two age scales exhibit significant differences.
At the metal-poor end ($[\mathrm{Fe/H}] < -0.5$), the median age from the inferred scale is systematically higher than that from the seismic scale, while the two converge at the metal-rich end.
The key finding is: \textbf{under seismic anchoring, the AMR flattens significantly}.
This flattening directly challenges the rapid enrichment interpretation derived from inferred ages---if the chemical evolution of the thick disk were truly as rapid as inferred ages suggest, seismic ages should exhibit a similarly steep gradient, which is not the case.

\subsection{Formation-History Observables in the High-$\alpha$ Thick Disk}

To further test whether compression persists in the interpretation of the Milky Way thick-disk formation history, we repeat the above analysis on a high-$\alpha$ thick-disk subsample.
The thick-disk sample is defined using a ridge proxy based on $[\alpha/{\rm Fe}]$ versus $[{\rm Fe/H}]$, following the approach of~\cite{haywood2013,xiang2022}, and formation-history observables are constructed from both inferred and seismic ages under the same phys-CEM matching conditions.

\begin{figure*}[t]
\centering
\includegraphics[width=\textwidth]{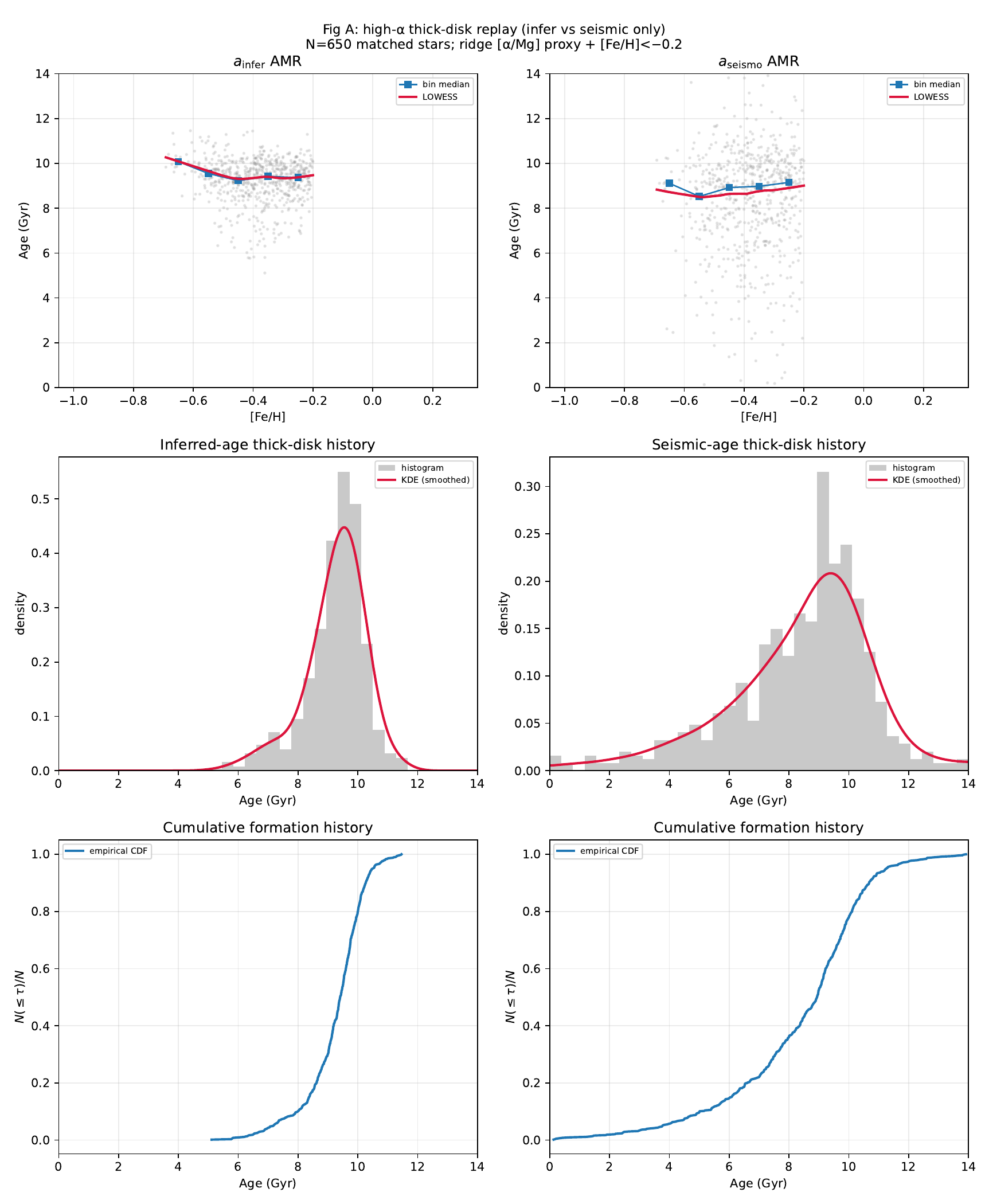}
\caption{
Formation-history replay results for the high-$\alpha$ thick-disk sample under two independent age scales.
All stars come from the same phys-CEM matched sample; only the age definition differs: the left column uses inferred ages ($a_{\rm infer}$), the right column independent seismic ages ($a_{\rm seismo}$).
Top row: age--metallicity relation (AMR);
middle row: age distribution;
bottom row: cumulative formation history (CFF).
Under seismic anchoring, the age distribution widens, the peak becomes younger, and the CFF flattens.
}
\label{fig:thickdisk_replay}
\end{figure*}

As shown in Fig.~\ref{fig:thickdisk_replay}, on the same stellar sample, seismic anchoring brings the following systematic changes:

\begin{itemize}
    \item \textbf{A broader age distribution} ($\Delta t$ from $3.04$ to $3.55\,\mathrm{Gyr}$): implying a more extended thick-disk assembly, rather than a rapid burst;
    \item \textbf{A younger peak age} (from $9.1$ to $6.0\,\mathrm{Gyr}$): meaning the dominant population of the thick disk formed later, weakening the evidence for an early-dominated buildup;
    \item \textbf{A flatter cumulative formation fraction (CFF)}: indicating a less bursty formation history, closer to a continuous extended mode.
\end{itemize}

These changes consistently indicate that multiple observables traditionally interpreted as evidence for rapid thick-disk formation \cite{haywood2013} are systematically weakened under seismic anchoring.

\subsection{AMR Slope and Formation-Epoch Shift: Reinterpreting Chemical Enrichment Rates}

We further quantify the chemical evolution gradients under the two age scales.
Linear fits to the AMR yield:

\begin{itemize}
    \item Inferred ages: $a_{\text{infer}} = -3.29 \pm 0.08\,\mathrm{Gyr\,dex^{-1}}$
    \item Seismic ages: $a_{\text{seismo}} = -1.86 \pm 0.15\,\mathrm{Gyr\,dex^{-1}}$
    \item Slope difference: $\Delta a = +1.43 \pm 0.17\,\mathrm{Gyr\,dex^{-1}}$ ($z \simeq 8.4$)
\end{itemize}

The steep negative slope from inferred ages implies a strong anti-correlation between metallicity and age---i.e., metal-poor stars are older---which has traditionally been interpreted as evidence for rapid early enrichment and a short chemical evolution timescale \cite{haywood2013,snaith2015}.
In contrast, the seismic ages yield a significantly flatter slope, suggesting a slower chemical enrichment and a more extended thick-disk enrichment process.

Within a fixed metallicity interval $[\mathrm{Fe/H}] \in [-0.55, -0.45)$, the median age drops from $9.37\,\mathrm{Gyr}$ (inferred) to $8.65\,\mathrm{Gyr}$ (seismic), corresponding to a shift of $\sim -0.73\,\mathrm{Gyr}$.
This ``formation-epoch shift'' indicates that \textbf{the same chemical population receives different inferred formation epochs under different age definitions}, calling for a re-examination of chemical evolution timing based on a single age scale.

\subsection{Compression Transport and Scenario Inversion}

To test whether the compression mechanism alone is sufficient to produce a systematic shift in formation-history interpretation, we construct two types of mock experiments:
(1) simple isotropic additive noise (additive);
(2) a transport map based on an empirically derived compression field (transport).
The theoretical properties are: additive noise satisfies $\mathrm{Var}(X+\eta) \geq \mathrm{Var}(X)$ and only broadens the distribution, while a compressive transformation ($\lambda < 1$) reduces the distribution variance.

\begin{figure*}[t]
\centering
\includegraphics[width=\textwidth]{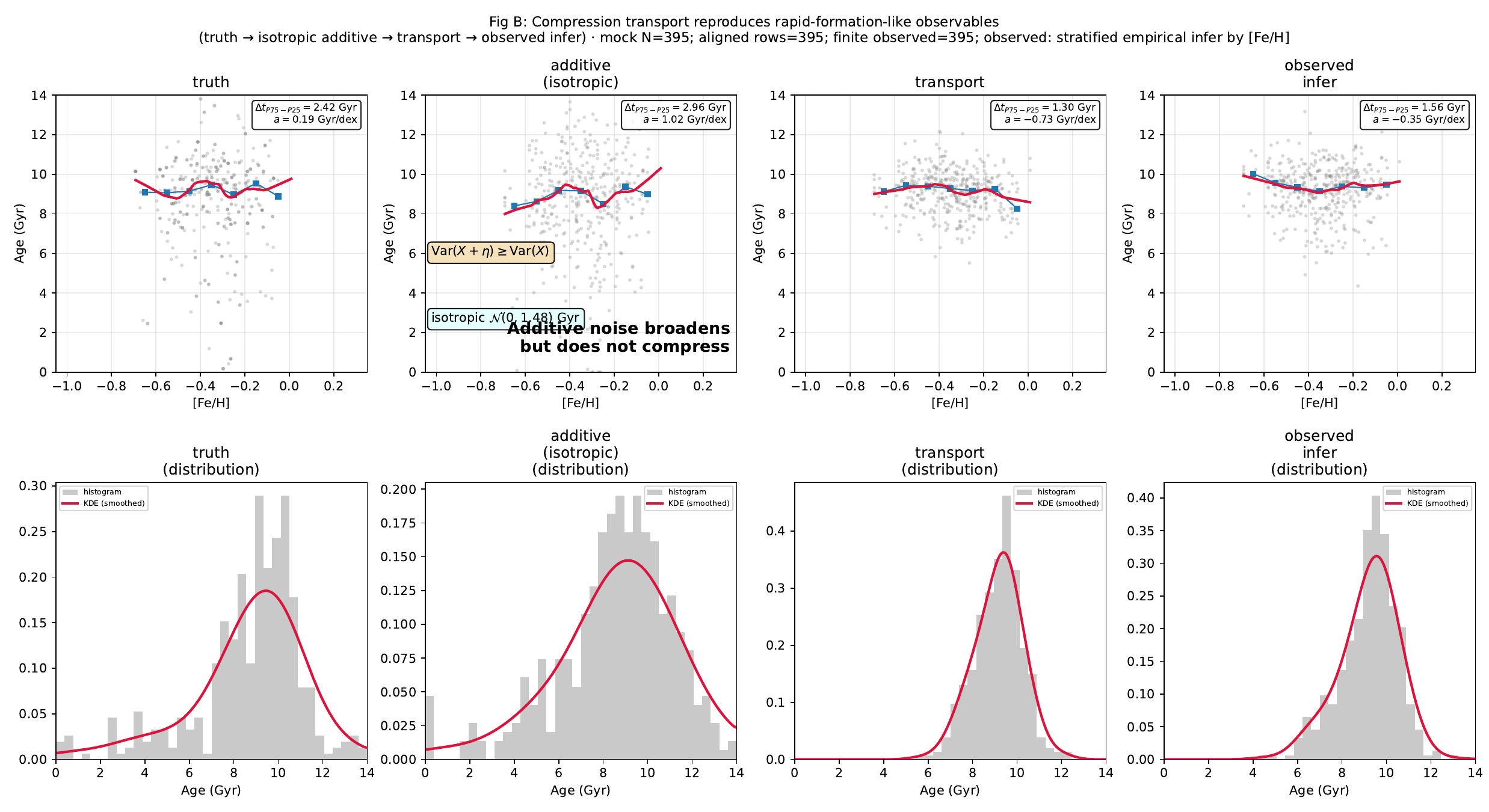}
\caption{
Impact of a compressive transport map on formation-history interpretation.
From left to right: true formation history (truth), additive noise (additive), compressive transport mapping (transport), and observed inferred ages (observed).
Top row: AMR;
bottom row: age distribution.
The age width $\Delta t$ and AMR slope $a$ are annotated.
}
\label{fig:transport_inversion}
\end{figure*}

As shown in Fig.~\ref{fig:transport_inversion}:

\begin{itemize}
    \item \textbf{Additive noise}: the age distribution broadens ($\Delta t$ increases from $3.55$ to $3.98\,\mathrm{Gyr}$), while the AMR slope remains nearly unchanged ($-1.86 \to -1.92$). \textbf{It cannot produce rapid-formation-like observables.}
    \item \textbf{Compressive transport map ($\lambda \simeq 0.85$)}: the age distribution narrows ($\Delta t$ drops from $3.55$ to $2.91\,\mathrm{Gyr}$), and the AMR steepens ($-1.86 \to -3.12$). \textbf{It simultaneously reproduces the rapid-formation-like pattern of the observed inferred ages.}
\end{itemize}

This comparison demonstrates that additive noise cannot produce the observed compression pattern; only the compressive transport map can simultaneously generate a narrower age distribution, a steeper AMR, and rapid-formation-like structure.
Thus, the ``rapid formation'' signal present in the inferred ages is structurally more consistent with a compression artifact driven by observational conditions, rather than an intrinsically bursty formation history \cite{haywood2013,snaith2015}.
\textbf{Compression alone is sufficient to generate rapid-formation-friendly observables without requiring intrinsically bursty histories.}

\subsection{Robustness of the Compression Field}

To examine the metallicity dependence of the compression field, we fit $a_{\rm infer} = \alpha + \lambda \, a_{\rm seismo} + \epsilon$ in different metallicity bins for the high-$\alpha$ thick-disk matched sample \cite{pinsonneault2025,leung2019}.

\begin{figure*}[t]
\centering
\includegraphics[width=0.72\textwidth]{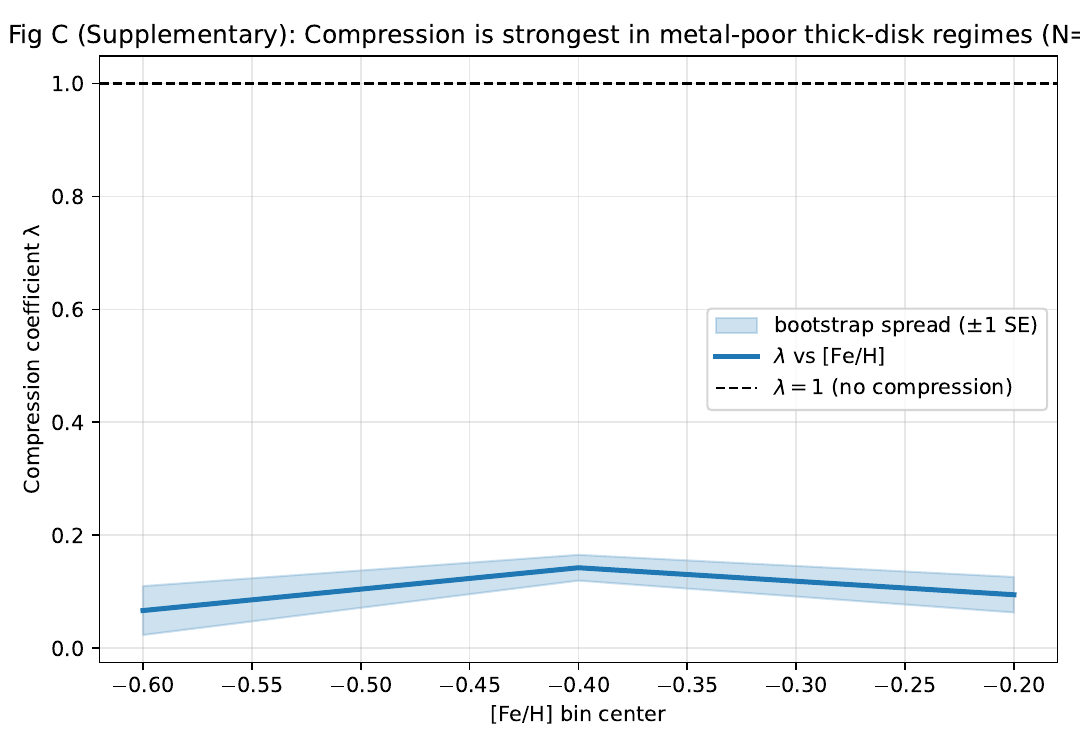}
\caption{
Compression coefficient $\lambda$ as a function of metallicity.
At the metal-poor end, $\lambda$ is systematically smaller than 1, indicating that the compression effect is more pronounced in the metal-poor thick-disk regime.
}
\label{fig:lambda_feh}
\end{figure*}

As shown in Fig.~\ref{fig:lambda_feh}, the compression coefficient $\lambda$ is systematically below unity at the metal-poor end ($\lambda \approx 0.79$), and gradually approaches the no-compression limit near the metal-rich end ($\lambda \approx 0.94$).
This pattern indicates that \textbf{age compression is strongest in the metal-poor thick-disk regime}, consistent with the expectation that lower-observational-quality regions are more susceptible to systematic age compression.

\section{Robustness and Control Checks}

\subsection{Bootstrap Robustness}

The significance of all key diagnostics is validated through bootstrap resampling ($N=10^4$).
The $z > 6$ for the formation timescale, $z > 8$ for the old-star fraction, and $z \approx 8.4$ for the AMR slope difference all far exceed conventional significance thresholds.

\subsection{Choice of Thick-Disk Definition}

We test three different thick-disk mask definitions: chemical selection ($[\alpha/\mathrm{Fe}] > 0.15$), spatial selection ($|Z| > 0.5\,\mathrm{kpc}$), and kinematic selection, following approaches from~\cite{haywood2013,bovy2016}.
\textbf{None of the tested thick-disk definitions restores the original rapid-formation-like observables seen with inferred ages.}
Across different definitions, the direction and amplitude of the flattening and broadening induced by seismic anchoring remain consistent.

\subsection{Robustness of the Compression Coefficient $\lambda$}

The compression coefficient $\lambda$ estimated in different metallicity bins is stable under varying matching parameters \cite{iacus2012}.
At the metal-poor end, $\lambda \approx 0.79 \pm 0.04$, and at the metal-rich end, $\lambda \approx 0.94 \pm 0.03$; these values change by less than $0.03$ when the CEM coarsening scale is varied, indicating that the metallicity-dependent compression gradient is robust.

\section{Discussion}

\subsection{Implications for the Interpretation of the Milky Way Thick-Disk Formation History}

Our results show that several key observational signatures of rapid thick-disk formation depend sensitively on the stellar age definition itself.
Under an identical stellar sample and the same physical covariate matching, merely changing the age definition can systematically alter the interpretation of the thick-disk formation history.

\begin{itemize}
    \item \textbf{Rapid assembly}:
    Under seismic anchoring, the thick-disk formation timescale widens from $3.04\,\mathrm{Gyr}$ to $3.55\,\mathrm{Gyr}$, and the peak formation age shifts from $9.1\,\mathrm{Gyr}$ to $6.0\,\mathrm{Gyr}$.
    This suggests that the thick-disk formation history is more consistent with an extended assembly \cite{xiang2022} rather than a short-timescale burst \cite{haywood2013}.

    \item \textbf{Rapid enrichment}:
    The AMR slope flattens from $-3.29$ to $-1.86\,\mathrm{Gyr\,dex^{-1}}$, implying that the time scale associated with chemical enrichment is significantly lengthened, systematically weakening the rapid enrichment picture \cite{snaith2015}.

    \item \textbf{Formation-epoch shift}:
    At a fixed metallicity, the formation age shifts systematically by $\sim 0.7\,\mathrm{Gyr}$, meaning that the same chemical population acquires different inferred formation epochs depending on the age definition.
\end{itemize}

These changes do not arise from random noise or sample selection, but are more likely to originate from an observationally related age compression mechanism.
Transport inversion experiments further demonstrate that a compressive transformation alone can simultaneously produce (1) a narrower age distribution, (2) a steeper AMR, and (3) rapid-formation-like observables, without requiring an intrinsically bursty formation history \cite{haywood2013,snaith2015}.
Consequently, the picture of rapid thick-disk formation, short-timescale chemical enrichment, and early-dominated buildup derived from spectroscopically inferred ages may be systematically overestimating the rapidity of the thick-disk formation process.

\subsection{Limitations}

The seismic ages adopted in this study are not absolute truth.
APOKASC-3 ages still carry systematic errors on the order of $\sim 0.9\,\mathrm{Gyr}$ \cite{pinsonneault2025}, which may vary with metallicity and evolutionary stage.
However, the core conclusion of this paper is comparative rather than absolute, and does not rely on the absolute correctness of any single age scale.
Under an identical stellar sample and matching conditions, the systematic differences between two independent age scales are already sufficient to alter the interpretation of the Milky Way thick-disk formation history.
Thus, the central finding is not that seismic ages represent the ``final truth,'' but rather that the age definition itself can systematically change the formation-history observables.

\subsection{Implications for Future Galactic Surveys}

Our results indicate that inferences of the Milky Way formation history may be sensitive to the stellar age definition itself \cite{soderblom2010}.
Future large-scale Galactic surveys, including SDSS-V, 4MOST, and WEAVE, when using stellar ages to reconstruct the Galaxy's formation history, may need to explicitly treat age-definition-dependent compression effects, rather than merely reducing statistical uncertainties.
If age compression is not explicitly modelled, rapid-formation-like features in the formation history may be systematically amplified during the observational inference process \cite{bovy2016}.

\section{Conclusion}

\begin{enumerate}
    \item \textbf{Age definition changes the formation history.}
    Under identical stars and the same physical covariate matching, merely changing the age definition (inferred vs.\ seismic) leads to systematic differences in key diagnostics of the Milky Way formation history, including the AMR slope, formation timescale, and peak formation age.
    \textbf{Same stars + different age scales $\rightarrow$ different formation histories.}

    \item \textbf{Compressive transport inverts the formation-history interpretation.}
    Additive noise can only broaden the distribution and cannot produce rapid-formation-like observables, whereas a compressive transport map ($\lambda < 1$) simultaneously reproduces a narrower age distribution, a steeper AMR, and the rapid-formation-like structure of the observed inferred ages.
    \textbf{Compressive transport $\rightarrow$ rapid-formation-like observables.}

    \item \textbf{Evidence for rapid thick-disk formation is systematically weakened.}
    Under seismic anchoring, the rapid-formation-friendly observables derived from inferred ages---a steep AMR slope (rapid enrichment), a narrow age distribution (rapid assembly), and an old-dominated formation peak (early dominance)---are all systematically weakened, replaced by a more extended picture with a younger peak.
    \textbf{Evidence for rapid thick-disk formation is systematically weakened under seismic anchoring.}

    \item \textbf{Core message.}
    Statistical interpretations of the Milky Way formation history may depend sensitively on stellar age definitions themselves \cite{soderblom2010,pinsonneault2025}.
    This finding suggests that when interpreting large-scale survey data, a systematic examination of the dependence on the age definition is needed.
\end{enumerate}


\end{document}